\documentclass[11pt,a4paper]{article}
\usepackage[T1]{fontenc}
\usepackage[utf8]{inputenc}
\usepackage{lmodern}
\usepackage[margin=2.5cm]{geometry}
\usepackage{amsmath,amssymb}
\usepackage{graphicx}
\usepackage[numbers,sort&compress]{natbib}
\usepackage{textcomp,microtype}
\usepackage[section,below]{placeins}
\usepackage{caption}
\usepackage[hidelinks]{hyperref}

\title{\bfseries MASON: Materials Atomistic Structure builder Operated in Natural language}
\author{\parbox{\dimexpr\textwidth-2\tabcolsep\relax}{\centering
Xinyue Zhang$^{1,\dagger}$, Jixiang Li$^{2,\dagger}$, Bin Shao$^{1,3,4,*}$,\\
Baishun Yang$^{5}$, Zhiyang Liu$^{1,3,4,*}$, Weichao Wang$^{1,4,6,7,*}$\\[0.8em]
\small
$^{1}$College of Electronic Information and Optical Engineering, Nankai University, Tianjin 300350, China\\[0.25em]
$^{2}$School of Physics, Nankai University, Tianjin 300071, China\\[0.25em]
$^{3}$Tianjin Key Laboratory of Optoelectronic Sensor and Sensing Network Technology, Nankai University, Tianjin 300350, China\\[0.25em]
$^{4}$Shenzhen Research Institute of Nankai University, Shenzhen 518057, China\\[0.25em]
$^{5}$CIC nanoGUNE BRTA, Tolosa Hiribidea 76, San Sebasti\'an 20018, Spain\\[0.25em]
$^{6}$College of Energy and Environment Science, Yunnan Normal University, Kunming 650500, China\\[0.25em]
$^{7}$Southwest United Graduate School, Kunming 650092, China\\[0.7em]
\footnotesize
$^{\dagger}$These authors contributed equally to this work.\\
$^{*}$Corresponding authors: \href{mailto:bshao@nankai.edu.cn}{bshao@nankai.edu.cn};
\href{mailto:liuzhiyang@nankai.edu.cn}{liuzhiyang@nankai.edu.cn};
\href{mailto:weichaowang@nankai.edu.cn}{weichaowang@nankai.edu.cn}
}}
\date{}
\begin{document}
\maketitle

\begin{abstract}
Building reliable atomistic models is a critical and often challenging step in first-principles calculations. Although large language model agents can assist with structure generation, they lack direct capabilities for crystallographic operations and therefore struggle to construct complex structures reliably. Here we present MASON, an AI-native toolkit for atomistic structure construction that can be invoked by agents through the Model Context Protocol (MCP). In an end-to-end example, an agent builds a CO/Pt(111) adsorption model from a natural language request. By equipping agents with dedicated structure-building tools, MASON enables reliable and efficient atomistic model construction and provides a foundation for autonomous first-principles workflows.

\end{abstract}


\section{Introduction}
Language-model agents are beginning to move beyond conversation into scientific research. Recent studies have demonstrated their use in hypothesis generation \cite{gottweis2026coscientist}, molecular design \cite{swanson2025virtuallab}, chemical synthesis \cite{boiko2023autonomous,bran2024chemcrow} and atomistic simulation \cite{ghafarollahi2025atomagents} across biology, chemistry and materials science, and even in the automation of a complete research cycle from idea to manuscript \cite{lu2026aiscientist}. A common feature of these systems is their reliance on domain-specific tools that connect language-model reasoning with scientific operations, following the tool-use paradigm in which a model combines reasoning with calls to external functions \cite{yao2023react,schick2023toolformer}. Such tools are therefore central to extending agents into scientific practice \cite{ramos2025review,lei2024perspective}.

In first-principles calculations \cite{hohenberg1964inhomogeneous,kohn1965selfconsistent}, constructing a reliable atomistic structure is a critical first step. Mature libraries such as pymatgen \cite{ong2013pymatgen} and ASE \cite{larsen2017ase} provide powerful capabilities for tasks including defect construction \cite{freysoldt2014defects,broberg2018pycdt,kavanagh2024doped}, surface generation \cite{sun2013slabs} and heterostructure matching \cite{zur1984lattice,ding2016epitaxial}, and graphical programs such as VESTA \cite{momma2011vesta} serve interactive model building. However, these tools were designed for researchers, either as Python APIs or as interactive programs, rather than as tools that agents can invoke directly. Agent systems for materials science have so far used language models mainly to query materials databases \cite{chiang2024llamp} or to drive material-specific generation pipelines \cite{kang2024chatmof}; general crystallographic operations still have to be expressed as code. Agents must therefore translate scientific intent into task-specific code, introducing an additional and error-prone layer between reasoning and structure generation. Because errors in the initial structure propagate through all subsequent calculations, reliable, agent-oriented tools for atomistic structure construction are particularly important.

To address this gap, we present Materials Atomistic Structure builder Operated in Natural language (MASON), implemented as a Model Context Protocol (MCP) server \cite{mcp2024spec}. MASON provides 39 atomistic structure tools that agents can invoke directly, covering structure parsing and validation, supercell and defect construction, surface and adsorbate generation, and heterostructure assembly. It enables an agent to translate a research goal expressed in natural language into a sequence of structure-building operations and execute them through dedicated tools rather than ad hoc code. This provides a more reliable and efficient route to atomistic structure construction and establishes a foundation for autonomous first-principles workflows.
\section{Overview of MASON}
\label{sec:overview}

\subsection{Architecture}

\begin{figure}[!htb]
  \centering
  \includegraphics[width=\linewidth]{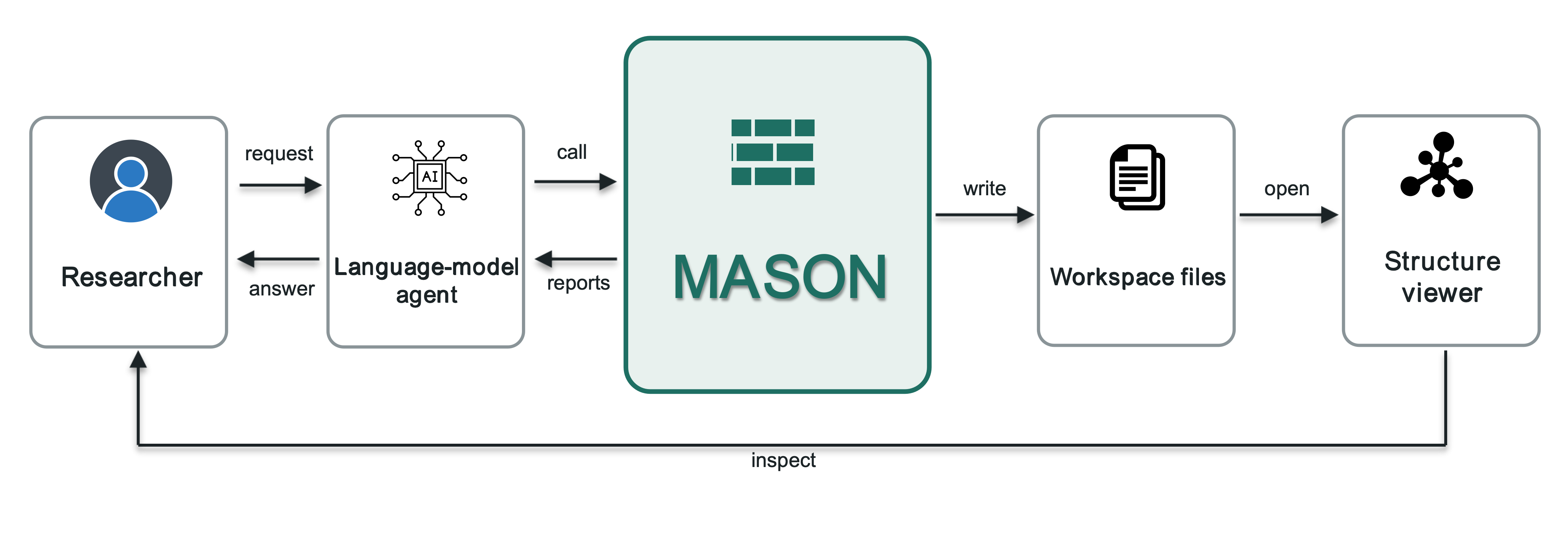}
  \caption{Architecture of MASON. The researcher's request goes to the
  language-model agent, which calls MASON through the Model Context Protocol.
  MASON builds the structure, reports back to the agent and writes the result to
  the workspace, where the structure viewer opens it for inspection.}
  \label{fig:arch}
\end{figure}

MASON separates workflow planning from crystallographic operations (Fig.~\ref{fig:arch}). The researcher
states a goal in natural language. The language-model agent decomposes the goal
into a sequence of tool calls and passes each call to MASON through the Model
Context Protocol. MASON performs the crystallographic operation, writes the
result to the workspace as a structure file together with an interactive
three-dimensional page, and returns a report of what it has built. The agent
reads this report before planning the next call, so that each step is based on
the structure actually built. The structure viewer opens each file as it is
written, and the researcher can inspect the result at any point without leaving
the session.

\begin{figure}[!tb]
  \centering
  \includegraphics[width=\linewidth,height=0.74\textheight,keepaspectratio]{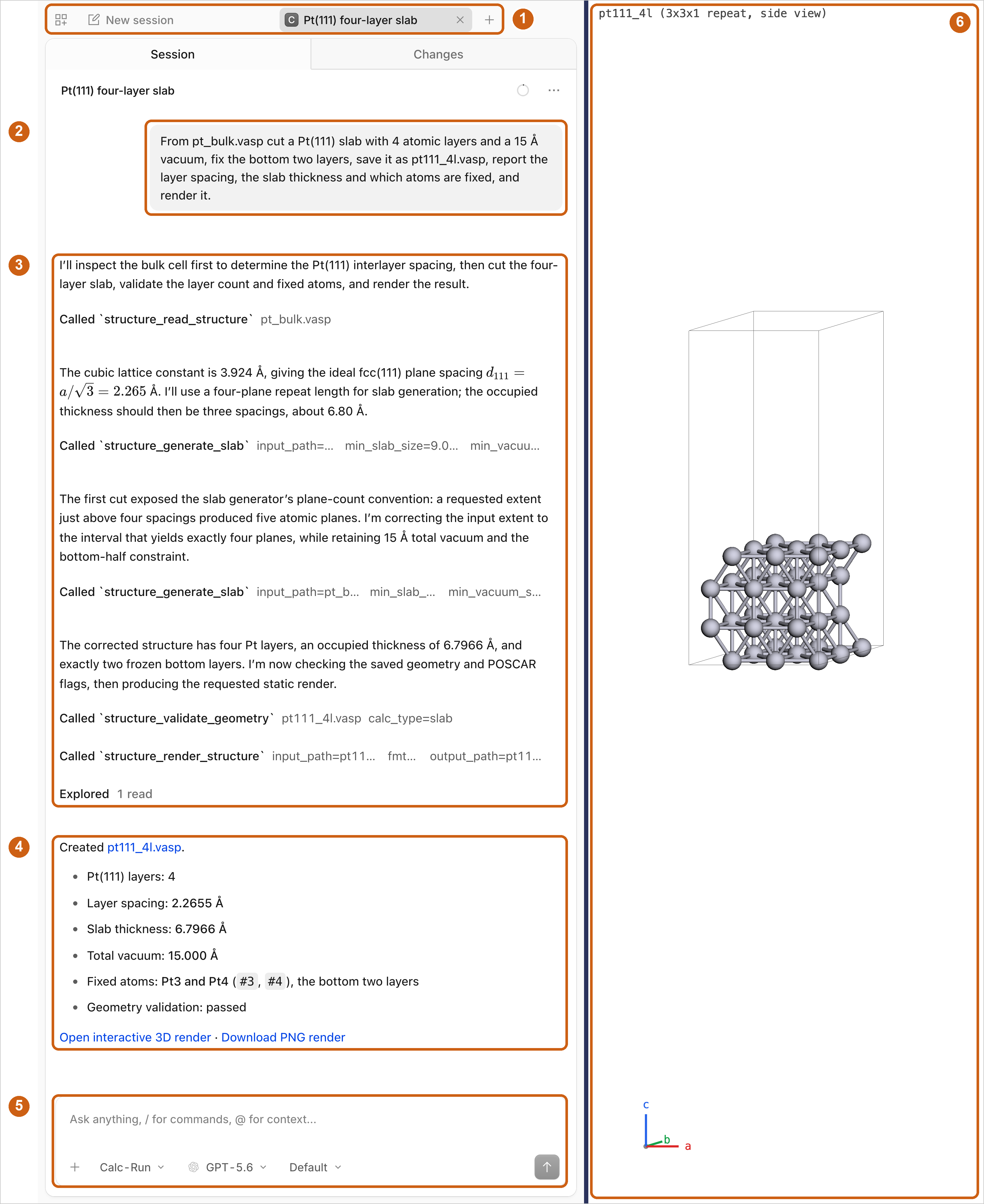}
  \caption{The MASON interface during the construction of a four-layer Pt(111)
  slab. In the conversation panel on the left, (1) session tabs, (2) the
  researcher's request, (3) the agent's reasoning together with the MASON tools
  it calls, in the order in which they are called, (4) the agent's summary of the
  resulting structure, and (5) the input box, in which the agent framework and
  the language model are selected. On the right, (6) the structure viewer,
  showing the slab from the side as a $3\times3\times1$ repetition.}
  \label{fig:workbench}
\end{figure}

The entire session is displayed in a single window (Fig.~\ref{fig:workbench}).
The conversation panel on the left shows the request, the agent's reasoning,
the tools it calls and the parameters it passes, in the order in which they are
called. The viewer on the right shows the resulting structure, which can be
rotated, zoomed and inspected atom by atom. The example in
Fig.~\ref{fig:workbench}, the construction of a four-layer Pt(111) slab
(Section~\ref{sec:slab}), shows how the agent uses the report. The first cut
produced five atomic planes instead of the four requested; the agent read the
layer count in the report, corrected the slab extent, cut again, and then
validated and rendered the slab. All of these steps are visible to the
researcher in real time.

\FloatBarrier
\subsection{Tool classes}
\begin{figure}[!htb]
  \centering
  \includegraphics[width=\linewidth,height=0.74\textheight,keepaspectratio]{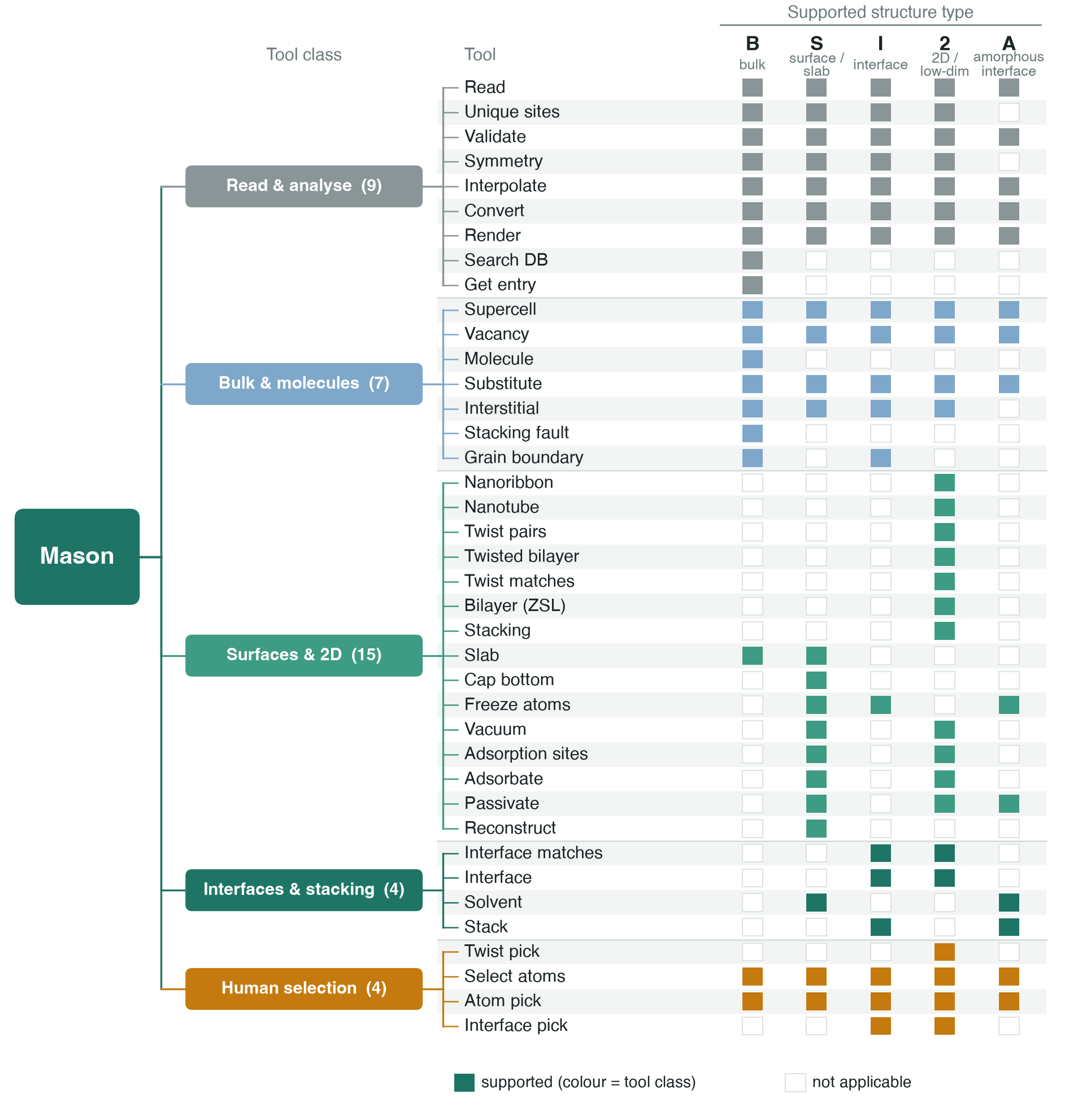}
  \caption{The MASON tool tree and the structure types accepted by each tool. Left, the five classes with the number of tools in each. Right, for every tool, a filled cell marks a supported structure type (B bulk, S surface or slab, I interface, 2 two-dimensional or low-dimensional, A amorphous interface) and an empty cell a type that does not apply.}
  \label{fig:tree}
\end{figure}

MASON provides 39 tools in five classes, organised according to the typical
structure-building workflow (Fig.~\ref{fig:tree}). The first class is for
structure analysis. It converts formats, determines the space group and the
inequivalent sites, checks the geometry, renders the structure and retrieves
entries from the Materials Project database
\cite{jain2013materials,ong2015api}. The second class handles bulk crystals and
molecules, including the construction of supercells, point defects, stacking
faults and grain boundaries. The third class covers surfaces and
low-dimensional structures, including slab generation, surface passivation,
adsorbate placement, and the construction of nanoribbons, nanotubes and twisted
bilayers. The fourth class constructs heterostructures: the lattice-matched
supercells of two materials are first identified, and the heterostructure is
then built from the selected match. The fifth class supports human-in-the-loop
selection. When a lattice match or a group of atoms should be chosen by the
researcher, the candidates are shown on an interactive page and the selection
is returned to the agent. All tools and the structure types they support are
listed in Fig.~\ref{fig:tree}, and the underlying algorithms are described in
Section~\ref{sec:functions}.

\FloatBarrier
\subsection{Design and implementation}

A language-model agent that builds a structure by itself writes ad hoc code for
every request, so the same request can give a different structure on each run,
and an error in the code is not visible in the file it produces. MASON replaces
code generation with tool calls. The crystallographic operations are
implemented in advance as tools, and the agent only selects the tool and its
parameters. The same call therefore always gives the same structure.

\begin{figure}[!htb]
  \centering
  \includegraphics[width=\linewidth]{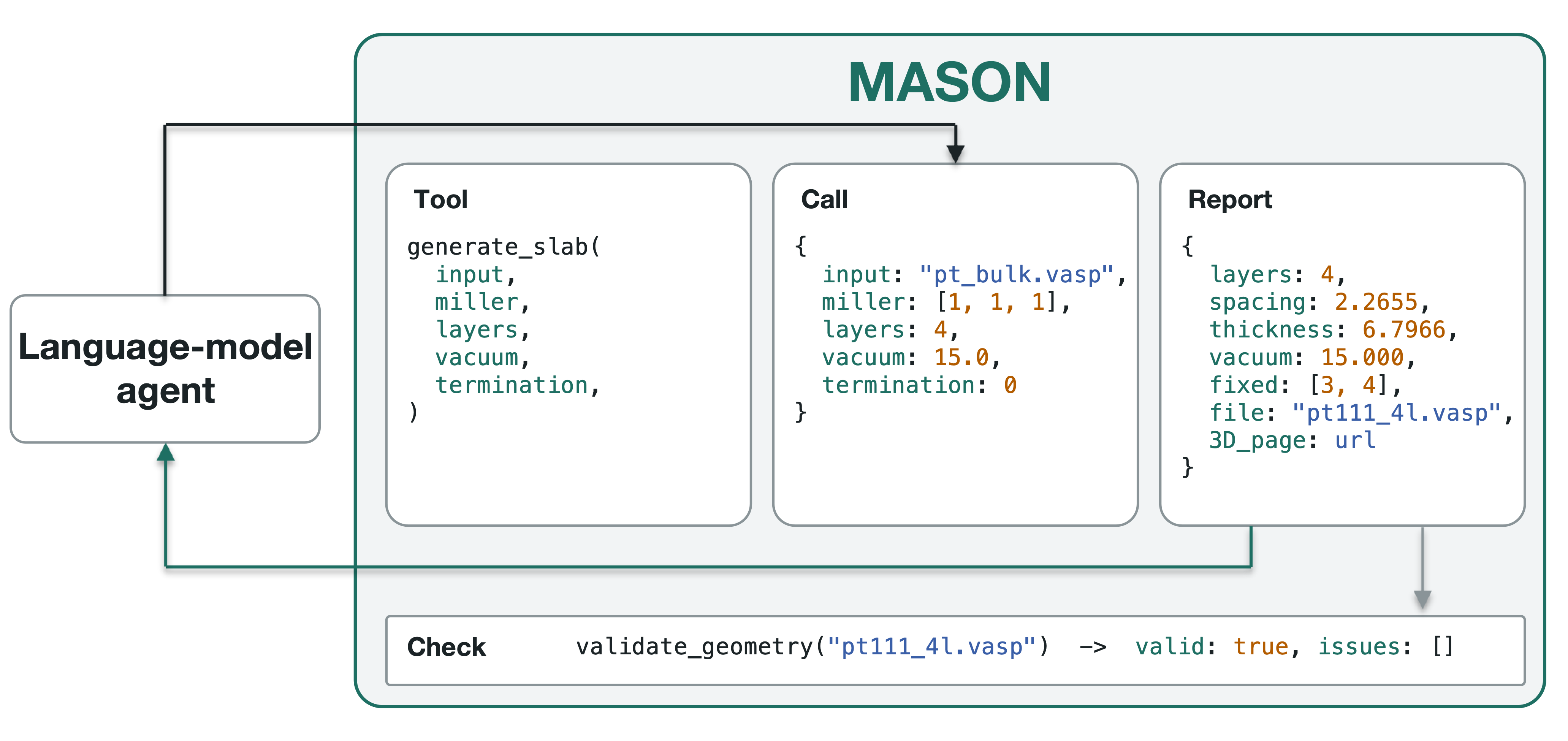}
  \caption{A MASON tool call, shown for the slab tool. The agent chooses the tool and fills in its parameters, MASON builds the structure and reports the layer spacing, thickness and other numbers of the slab it has built, and a second tool checks the geometry of the result.}
  \label{fig:anatomy}
\end{figure}

Fig.~\ref{fig:anatomy} illustrates a tool call, using the slab tool as an
example. The tool signature defines the available parameters, the call supplies
their values, and the report returns the quantities computed from the structure
that was built. Since the language model acts on the structure only through
such calls, an inappropriate parameter is revealed by the report and can be
corrected in the next call.

The crystallographic operations of MASON are built on pymatgen
\cite{ong2013pymatgen} and ASE \cite{larsen2017ase}, the symmetry analysis
uses spglib \cite{togo2024spglib}, and the three-dimensional pages are generated
with 3Dmol.js \cite{rego2015threedmol}. MASON is independent of the language
model and of the front end, and can be loaded as a server by any agent
framework that supports MCP. All tool outputs have been validated by comparison
with analytic values or with direct pymatgen and spglib calculations, and the
validations relevant to the examples are given in
Section~\ref{sec:functions}.
\FloatBarrier
\section{Capabilities and examples}
\label{sec:functions}

\FloatBarrier
\subsection{Structure analysis and validation}

The tools of this class read and analyse a structure file. The reading tool
returns the composition and the lattice parameters, the symmetry tool returns
the space group and the inequivalent sites, and the validation tool detects
overlapping atoms, degenerate cells and insufficient vacuum. These tools are
used both to characterise an initial structure and to verify the result of each
building step.

\begin{figure}[!tb]
  \centering
  \includegraphics[width=\linewidth,height=0.74\textheight,keepaspectratio]{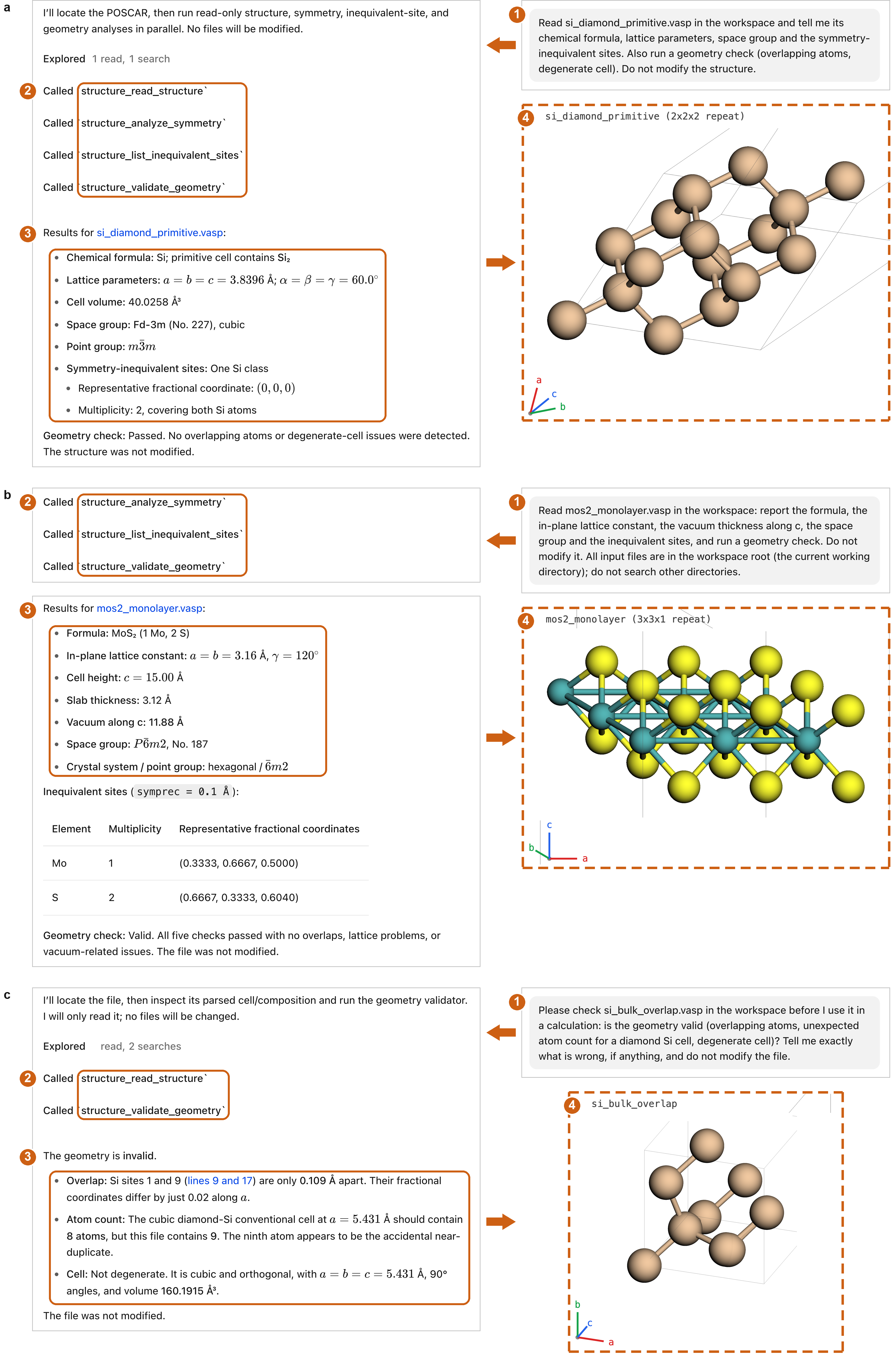}
  \caption{Reading, symmetry analysis and geometry validation. (a) The Si primitive cell, reading and symmetry analysis; (b) the MoS$_2$ monolayer, lattice, vacuum and symmetry; (c) validation of a Si cell that contains an overlapping atom. In each example (1) is the request, (2) the tools called, (3) the agent's summary and (4) the viewer, with the tool names and the key results boxed (here and in Figs.~\ref{fig:defect}--\ref{fig:interface}).}
  \label{fig:read}
\end{figure}

In the first example (Fig.~\ref{fig:read}a), the researcher requests the
formula, lattice parameters, space group and inequivalent sites of the Si
primitive cell, together with a geometry check. The agent calls the reading,
symmetry, inequivalent-site and validation tools in turn, and reports a
two-atom rhombohedral cell with $a = 3.84$~\AA{} and $\alpha = 60^\circ$, space
group $Fd\bar{3}m$ (No.~227), point group $m\bar{3}m$ and a single inequivalent
Si site of multiplicity 2. The structure passes the geometry check. The same
analysis of the MoS$_2$ monolayer (Fig.~\ref{fig:read}b) gives an in-plane
lattice constant of 3.16~\AA, space group $P\bar{6}m2$ (No.~187) and one
inequivalent site each for Mo and S, together with an S--Mo--S layer thickness
of 3.12~\AA{} and a vacuum of 11.88~\AA{} along the $c$ axis. The third example
(Fig.~\ref{fig:read}c) demonstrates error detection. The input file is a
conventional diamond-Si cell with an extra atom placed very close to an
existing one, and the researcher simply asks whether the file is suitable for a
calculation. The agent finds two Si atoms separated by only 0.11~\AA{} and an
atom count of 9 instead of 8, and reports that the structure is invalid.

\FloatBarrier
\subsection{Supercells and point defects}

The supercell tool expands a cell according to a given transformation matrix.
The point-defect tools create vacancies, substitutions and interstitials at
selected sites and report the resulting composition. Because a defect usually
lowers the symmetry, a symmetry analysis is normally performed afterwards, as
is common practice in defect calculations
\cite{broberg2018pycdt,kavanagh2024doped}.

\begin{figure}[!tb]
  \centering
  \includegraphics[width=\linewidth,height=0.74\textheight,keepaspectratio]{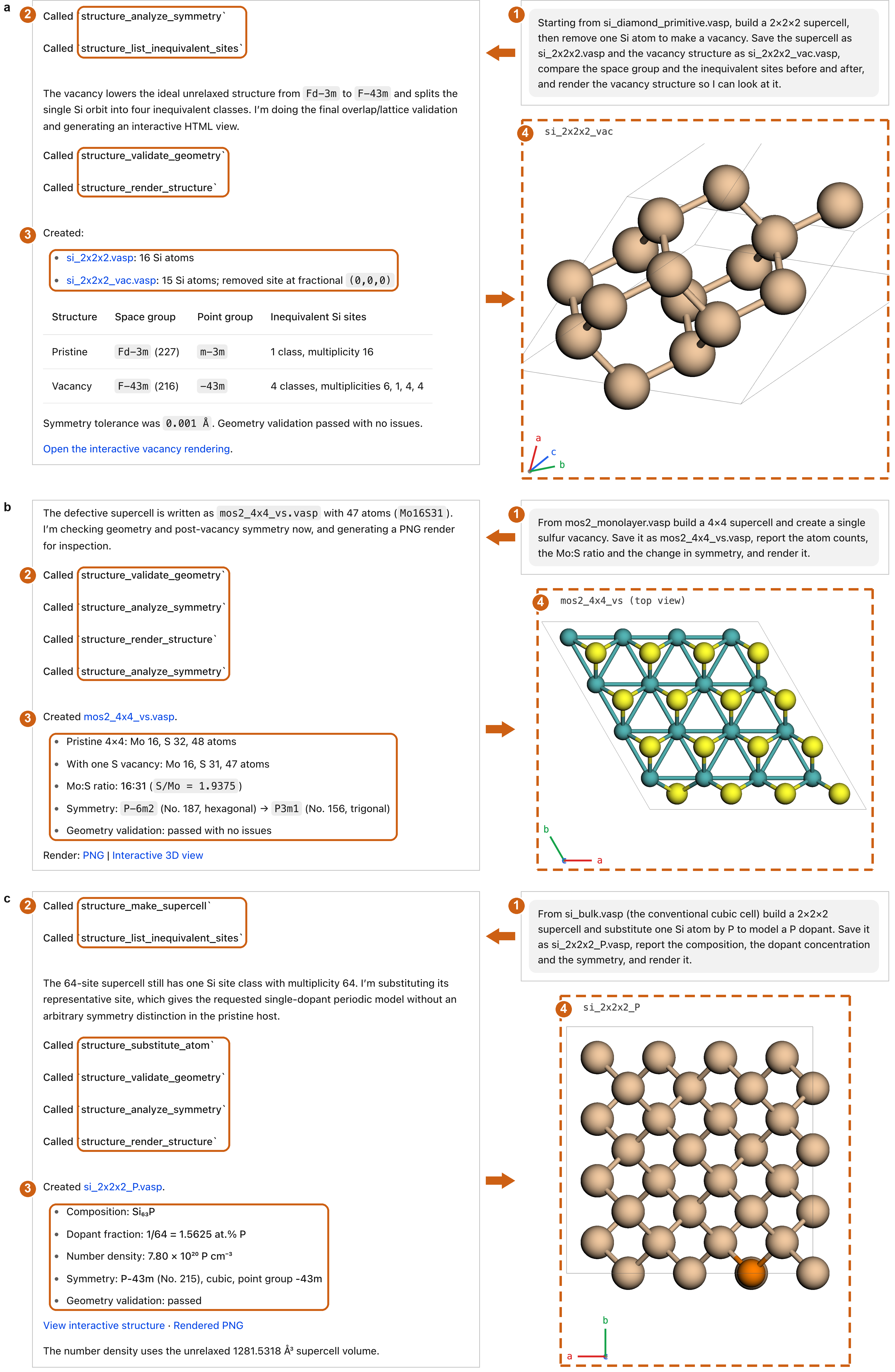}
  \caption{Supercells and point defects. (a) A $2\times2\times2$ supercell of the Si primitive cell with one Si vacancy; the summary tabulates the space group, point group and inequivalent sites before and after. (b) A sulfur vacancy in a $4\times4$ MoS$_2$ supercell (top view in the viewer). (c) A substitutional P dopant in the $2\times2\times2$ conventional Si supercell (the viewer shows the face on which the P atom, orange, is visible).}
  \label{fig:defect}
\end{figure}

In the first example (Fig.~\ref{fig:defect}a), the agent builds a
$2\times2\times2$ supercell of the Si primitive cell (16 atoms), removes the
atom at fractional coordinates $(0, 0, 0)$ to create a vacancy, and compares
the symmetry before and after. The pristine supercell retains $Fd\bar{3}m$ and
point group $m\bar{3}m$ with a single inequivalent site, whereas the vacancy
structure has $F\bar{4}3m$ (No.~216), point group $\bar{4}3m$ and four
inequivalent sites with multiplicities 6, 1, 4, and 4. The same vacancy in the
64-atom conventional supercell gives $P\bar{4}3m$ (No.~215) with nine
inequivalent sites. The two space groups share the point group $\bar{4}3m$ of
the vacancy site and differ only in the lattice type, because the periodic
images of the vacancy \cite{freysoldt2014defects} form an fcc array in the
first supercell and a simple cubic array in the second. Both results were
confirmed by direct spglib calculations. For the MoS$_2$ monolayer (Fig.~\ref{fig:defect}b), the agent
builds a $4\times4$ supercell and removes one S atom, and the summary gives a
report of the atom count ($48 \rightarrow 47$), the Mo:S ratio (16:31) and the
change of space group from $P\bar{6}m2$ to $P3m1$ (No.~156). In the third
example (Fig.~\ref{fig:defect}c), substituting one Si atom of the 64-atom
conventional supercell by P gives Si$_{63}$P, which corresponds to a dopant
concentration of 1.56~at.\% and has space group $P\bar{4}3m$. All three
structures pass the geometry check, and the defect sites can be inspected
directly in the viewer.

\FloatBarrier
\subsection{Slab models}
\label{sec:slab}

The slab tool cuts a surface of a given Miller index from a bulk structure,
following the slab-generation scheme of pymatgen \cite{sun2013slabs}, which
underlies the high-throughput surface-energy workflows built on it
\cite{tran2016surface}. The number of layers, the vacuum thickness, the
termination and the number of fixed bottom layers can be specified, and the
tool returns the geometric parameters of the resulting slab. For orientations
with more than one possible termination, all terminations are listed for
selection. A companion tool passivates the bottom surface with hydrogen to
saturate the dangling bonds of a one-sided slab model.

\begin{figure}[!tb]
  \centering
  \includegraphics[width=\linewidth,height=0.74\textheight,keepaspectratio]{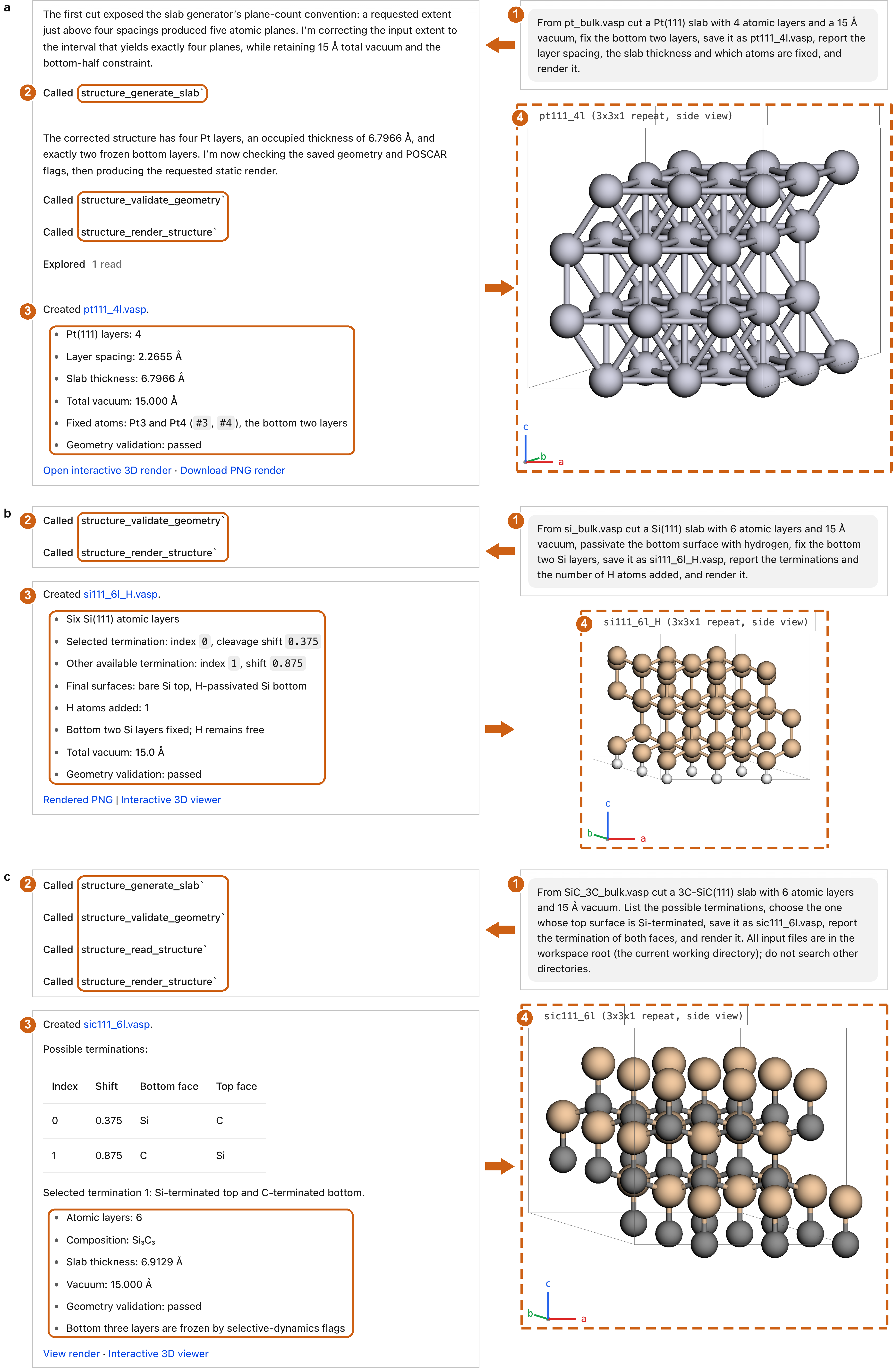}
  \caption{Surface slabs. (a) A four-layer Pt(111) slab with the bottom two layers fixed; (b) a six-layer Si(111) slab with a hydrogen-passivated bottom surface; (c) a six-layer 3C-SiC(111) slab, with both terminations listed and the Si-terminated top surface selected. The viewers show a $3\times3$ in-plane repetition from the side so that the layers can be counted.}
  \label{fig:slab}
\end{figure}

For a four-layer Pt(111) slab with 15~\AA{} of vacuum and the bottom two layers
fixed (Fig.~\ref{fig:slab}a), the agent reports an interlayer spacing of
2.27~\AA, in agreement with the analytic value $a/\sqrt{3}$ for
$a = 3.924$~\AA, a slab thickness of 6.80~\AA{} (three interlayer spacings) and
a vacuum of 15~\AA, and lists the atoms of the bottom two layers as fixed. For
a six-layer Si(111) slab cut from the conventional cell
(Fig.~\ref{fig:slab}b), the tool lists two terminations (cleavage shifts 0.375
and 0.875) and the agent selects the first. The bottom surface is passivated
with hydrogen, one H atom per surface Si atom, and the bottom two Si layers are
fixed while the H atoms are free to relax. The polar 3C-SiC(111) surface has
two terminations, with either C or Si on the top surface. Since a Si-terminated
top surface was requested, the agent selects the latter and obtains a six-layer
Si$_3$C$_3$ slab with a thickness of 6.91~\AA, 15~\AA{} of vacuum and the
bottom three layers fixed by default (Fig.~\ref{fig:slab}c). All three slabs
pass the geometry check.

\FloatBarrier
\subsection{Adsorption models}

Adsorption models are built with two tools. The site-finding tool identifies
the symmetry-inequivalent adsorption sites of a surface (top, bridge and
hollow), as in high-throughput adsorption workflows
\cite{montoya2017adsorption,boes2019catkit}, and distinguishes fcc from hcp
hollow sites according to the subsurface layers. The adsorbate tool places an
atom or a molecule at a selected site with a specified orientation and height,
and checks the minimum interatomic distance. Both tools report the distance
between periodic images of the adsorbate, which indicates the coverage
corresponding to the chosen surface cell.

\begin{figure}[!tb]
  \centering
  \includegraphics[width=\linewidth,height=0.74\textheight,keepaspectratio]{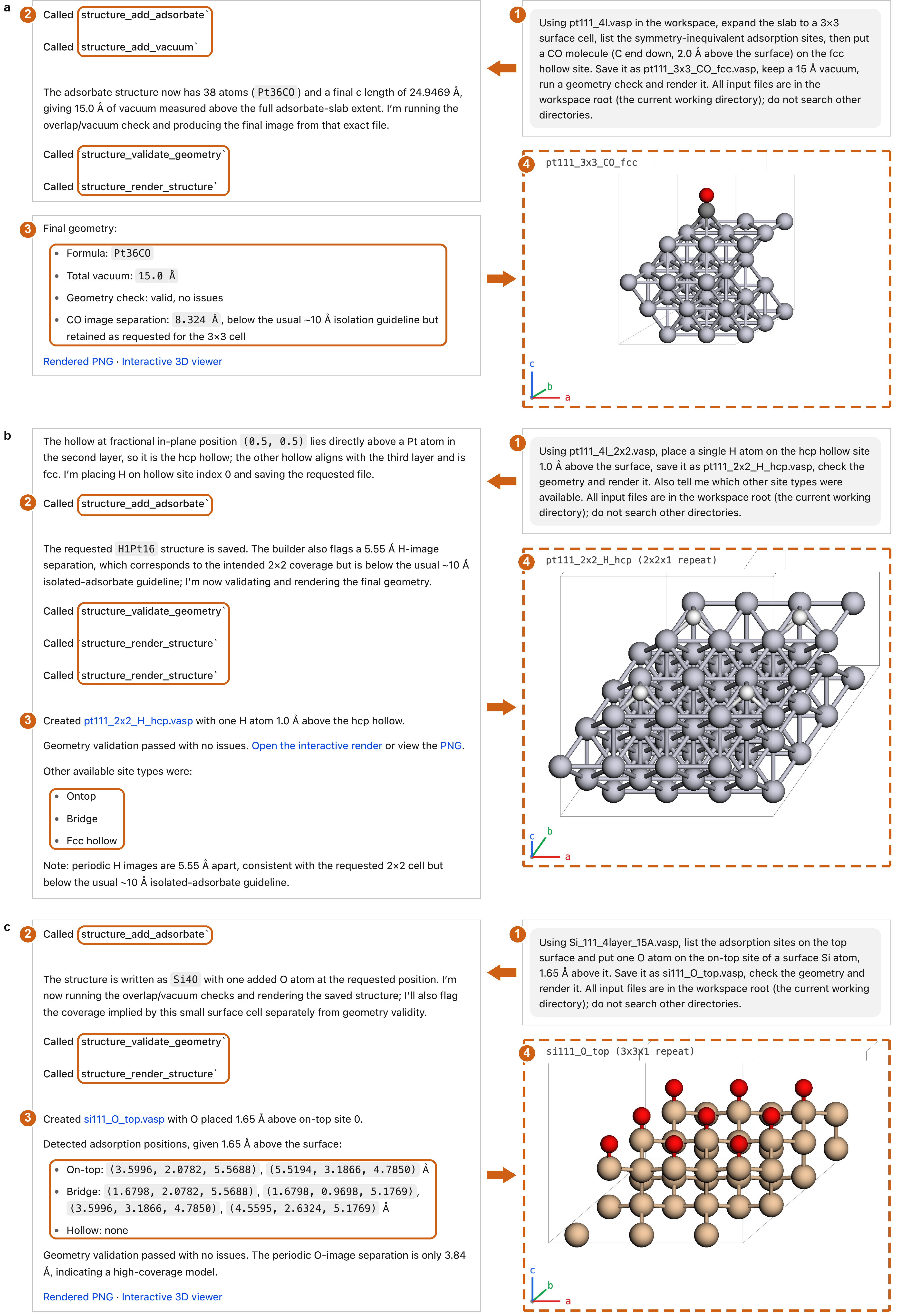}
  \caption{Adsorption sites and adsorbates. (a) CO, carbon end down, on the fcc hollow of a Pt(111)-$3\times3$ cell; (b) an H atom on the hcp hollow of Pt(111)-$2\times2$; (c) an O atom on the on-top site of Si(111). The viewers show the $3\times3$ cell itself (a) and $2\times2$ (b) and $3\times3$ (c) in-plane repetitions.}
  \label{fig:adsorb}
\end{figure}

In the first example (Fig.~\ref{fig:adsorb}a), the agent expands the four-layer
Pt(111) slab to a $3\times3$ surface cell, lists the inequivalent top, bridge,
hcp and fcc sites, and places a CO molecule, carbon end down, 2.0~\AA{} above
the fcc hollow site while keeping the vacuum at 15~\AA. The adsorption site is
specified in the request. The fcc site is used here because its stability
relative to the top site for CO on Pt(111) is a long-standing test case for
density-functional theory (DFT) calculations \cite{feibelman2001co}, and any other site can
be requested in the same way. The resulting Pt$_{36}$CO cell corresponds to a
coverage of 1/9 monolayer (ML), and the agent notes that the CO molecules in
neighbouring cells are 8.32~\AA{} apart. On a $2\times2$ cell of the same
surface, the agent places an H atom 1.0~\AA{} above the hcp hollow site and
lists the other available site types (Fig.~\ref{fig:adsorb}b). For Si(111), the
site-finding tool returns two top and four bridge sites. The agent places an O
atom 1.65~\AA{} above the topmost Si atom and notes that the O atoms in
neighbouring $1\times1$ cells are only 3.84~\AA{} apart, which corresponds to a
coverage of 1~ML (Fig.~\ref{fig:adsorb}c). All three structures pass the
geometry check.

\FloatBarrier
\subsection{Low-dimensional structures}

Two-dimensional materials and the nanostructures derived from them, such as
nanoribbons, nanotubes and twisted bilayers, are widely studied low-dimensional
systems \cite{novoselov2016vdw}. MASON builds these structures from a monolayer
with three tools. The nanoribbon tool cuts a periodic ribbon with a given edge
type and width, which together determine its electronic structure
\cite{son2006nanoribbons}. The nanotube tool rolls the monolayer into a tube
with chiral indices $(n, m)$ \cite{hamada1992nanotube}. The twisted-bilayer
tool lists the commensurate twist angles below a given number of atoms
\cite{lopesdossantos2007twist} and builds the bilayer for the selected angle.
Dangling bonds at the edges can be saturated with hydrogen using the
passivation tool.

\begin{figure}[!tb]
  \centering
  \includegraphics[width=\linewidth,height=0.74\textheight,keepaspectratio]{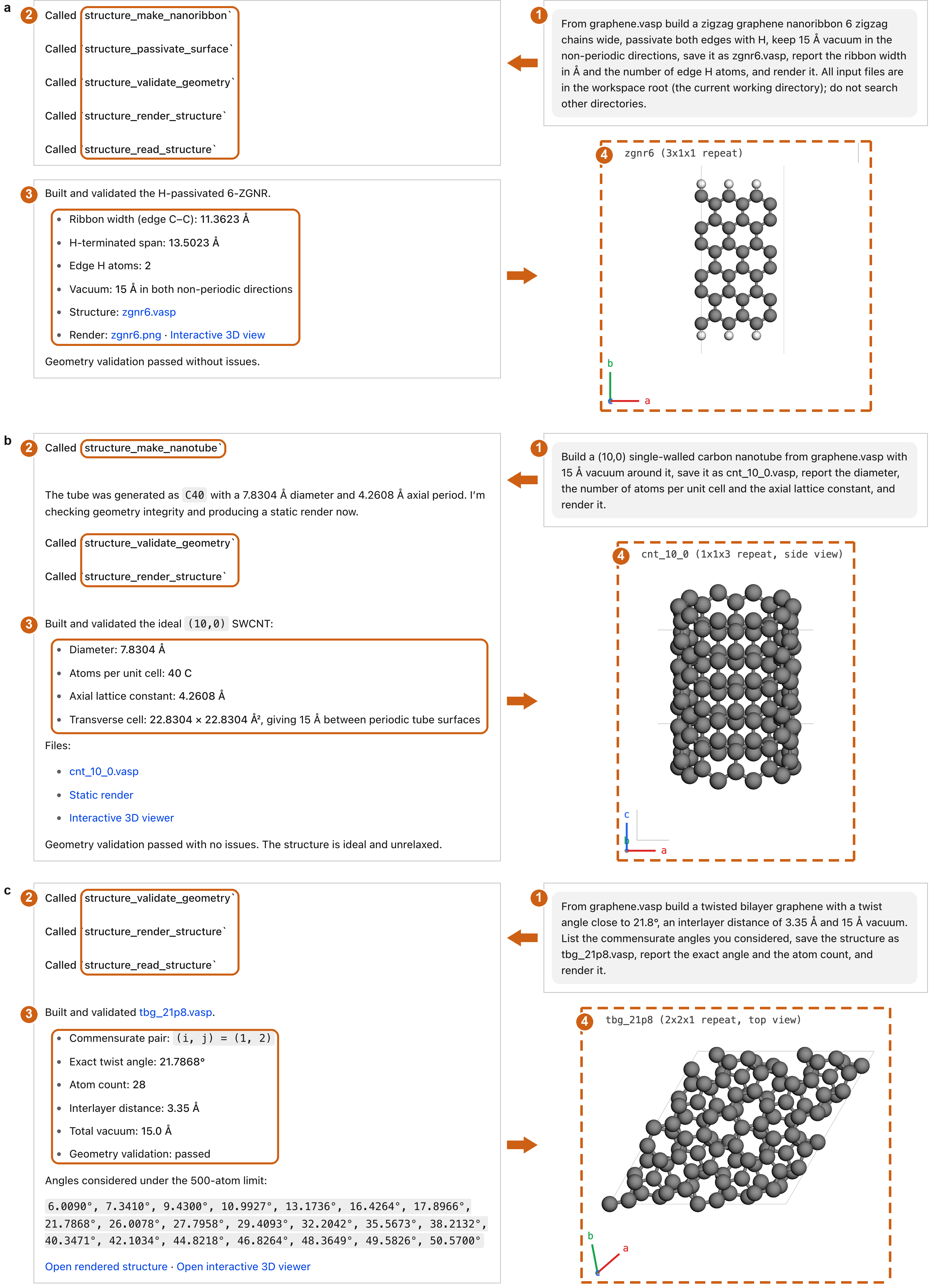}
  \caption{Low-dimensional structures. (a) A hydrogen-passivated zigzag graphene nanoribbon six zigzag chains wide (three periods along the ribbon axis in the viewer); (b) a (10,0) single-walled carbon nanotube (three periods, side view); (c) a commensurate twisted bilayer graphene at $21.79^\circ$ ($2\times2$ repetition, top view).}
  \label{fig:lowdim}
\end{figure}

For a hydrogen-passivated zigzag graphene nanoribbon with six zigzag chains
(Fig.~\ref{fig:lowdim}a), the agent reports a carbon-to-carbon width of
11.36~\AA, a width of 13.50~\AA{} including the hydrogen atoms, two edge H
atoms per period and 15~\AA{} of vacuum in both non-periodic directions. The
(10,0) carbon nanotube (Fig.~\ref{fig:lowdim}b) has a diameter of 7.83~\AA, 40
C atoms per period and an axial period of 4.26~\AA, and is placed in a
transverse cell of 22.83~\AA, which separates the walls of neighbouring tubes
by 15~\AA. For a twist angle close to $21.8^\circ$, the tool lists 21
commensurate angles below the 500-atom limit. The agent selects the closest
one, $21.79^\circ$ with commensurate indices (1, 2), and builds a 28-atom
bilayer with an interlayer distance of 3.35~\AA{} (Fig.~\ref{fig:lowdim}c). All
three structures pass the geometry check.

\FloatBarrier
\subsection{Heterostructures}

A heterostructure of two crystalline materials requires a common supercell in
which the lattice mismatch, and hence the strain, is acceptably small. The
matching tool searches for such lattice-matched supercells within given limits
of interface area and strain, using the algorithm of Zur and McGill
\cite{zur1984lattice}, which is implemented in pymatgen for substrate screening
\cite{ding2016epitaxial} and in MPInterfaces \cite{mathew2016mpinterfaces}. The
candidate matches are returned as a table and as an interactive selection page,
which list the interface area, the number of atoms and the von Mises strain of
each match. For a selected match, the construction tool builds the
heterostructure with a specified interlayer distance and vacuum thickness, and
applies the strain to one of the two materials, typically the film, so that it
fits the lattice of the other. The match can be selected by the researcher on
the interactive page or by the agent according to a specified criterion.

\begin{figure}[!tb]
  \centering
  \includegraphics[width=\linewidth,height=0.74\textheight,keepaspectratio]{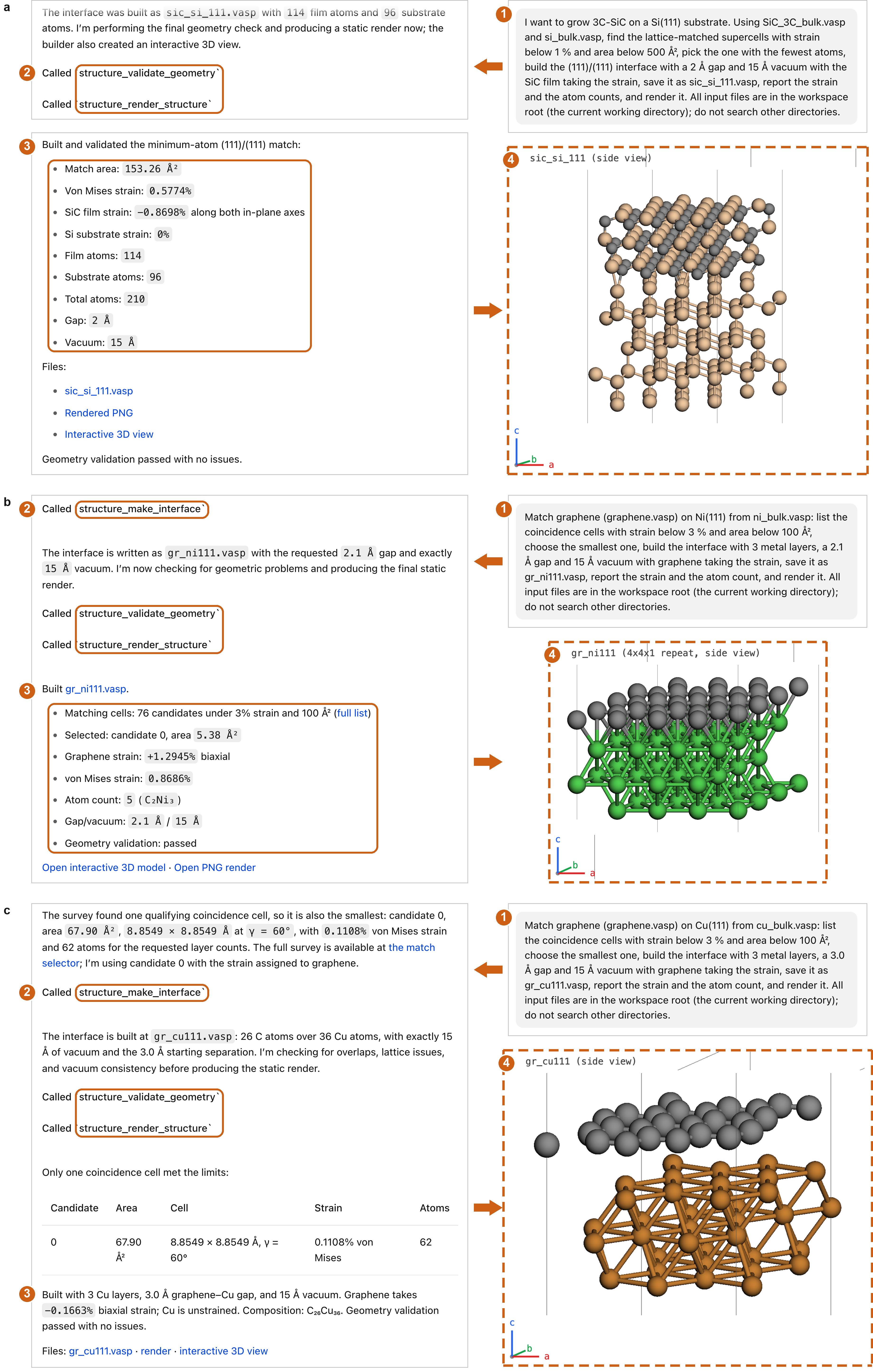}
  \caption{Lattice matching and heterostructure construction. (a) 3C-SiC(111) on Si(111) with a strain threshold of 1\% and an area limit of 500~\AA$^2$, the match with the fewest atoms selected; (b) graphene on Ni(111), a $1\times1$ graphene cell on a $1\times1$ Ni(111) cell ($4\times4$ repetition in the viewer); (c) graphene on Cu(111), a $\sqrt{13}\times\sqrt{13}$ graphene cell on a $\sqrt{12}\times\sqrt{12}$ Cu cell. The candidate counts and tables in the summaries are returned by the matching tool; all viewers are side views.}
  \label{fig:interface}
\end{figure}

The first example is 3C-SiC on Si(111), the classic heteroepitaxial system for
cubic SiC \cite{nishino1983sic} (Fig.~\ref{fig:interface}a). The researcher
specifies a maximum strain of 1\%, a maximum area of 500~\AA$^2$ and the match
with the fewest atoms as the selection rule. The matching tool returns twelve
candidates ranging from 210 to 638 atoms. The agent selects the 210-atom match,
which has an area of 153.26~\AA$^2$ and a von Mises strain of 0.58\%; the SiC
film is compressed by 0.87\% along both in-plane directions and the Si
substrate is unstrained. The agent then builds the heterostructure from 114 SiC
atoms and 96 Si atoms with an interlayer distance of 2~\AA{} and 15~\AA{} of
vacuum, and the structure passes the geometry check. Its in-plane lattice
constant, $a = b = 13.30$~\AA, equals $\sqrt{12}$ times the Si(111) surface
lattice constant of 3.84~\AA, as expected for a
$\sqrt{12}\times\sqrt{12}$ substrate supercell.

For graphene on Ni(111) (Fig.~\ref{fig:interface}b), the matching tool returns
76 candidates within limits of 3\% strain and 100~\AA$^2$. The agent selects
the smallest one, in which the $1\times1$ cell of graphene matches the
$1\times1$ cell of Ni(111), with an area of 5.38~\AA$^2$ and 5 atoms. This is
the registry observed experimentally for this system \cite{gamo1997graphene},
and it requires a biaxial tensile strain of 1.29\% in graphene (von Mises
strain 0.87\%). For graphene on Cu(111) \cite{gao2010graphene}, only one match
satisfies the same limits, a $\sqrt{13}\times\sqrt{13}$ graphene cell on a
$\sqrt{12}\times\sqrt{12}$ Cu(111) cell with an area of 67.90~\AA$^2$, 62
atoms, a von Mises strain of 0.11\% and a biaxial compressive strain of 0.17\%
in graphene (Fig.~\ref{fig:interface}c). For all three systems, the selected
match and its strain were reproduced by running the lattice-matching algorithm
of pymatgen directly with the same limits.

In all eighteen examples, the structures built by the agent pass the geometry
check, and the interlayer spacings, space groups and lattice matches are
consistent with analytic values or independent calculations.

\FloatBarrier
\section{End-to-end example: CO adsorption on Pt(111)}
\label{sec:workflow}

Each example above involves only one or two tools, whereas a model for a
practical calculation usually requires many. Here we show that the agent can
combine them within a single session, starting from one request in natural
language, to build a model that is ready for a DFT calculation.

\begin{figure}[!htb]
  \centering
  \includegraphics[width=\linewidth]{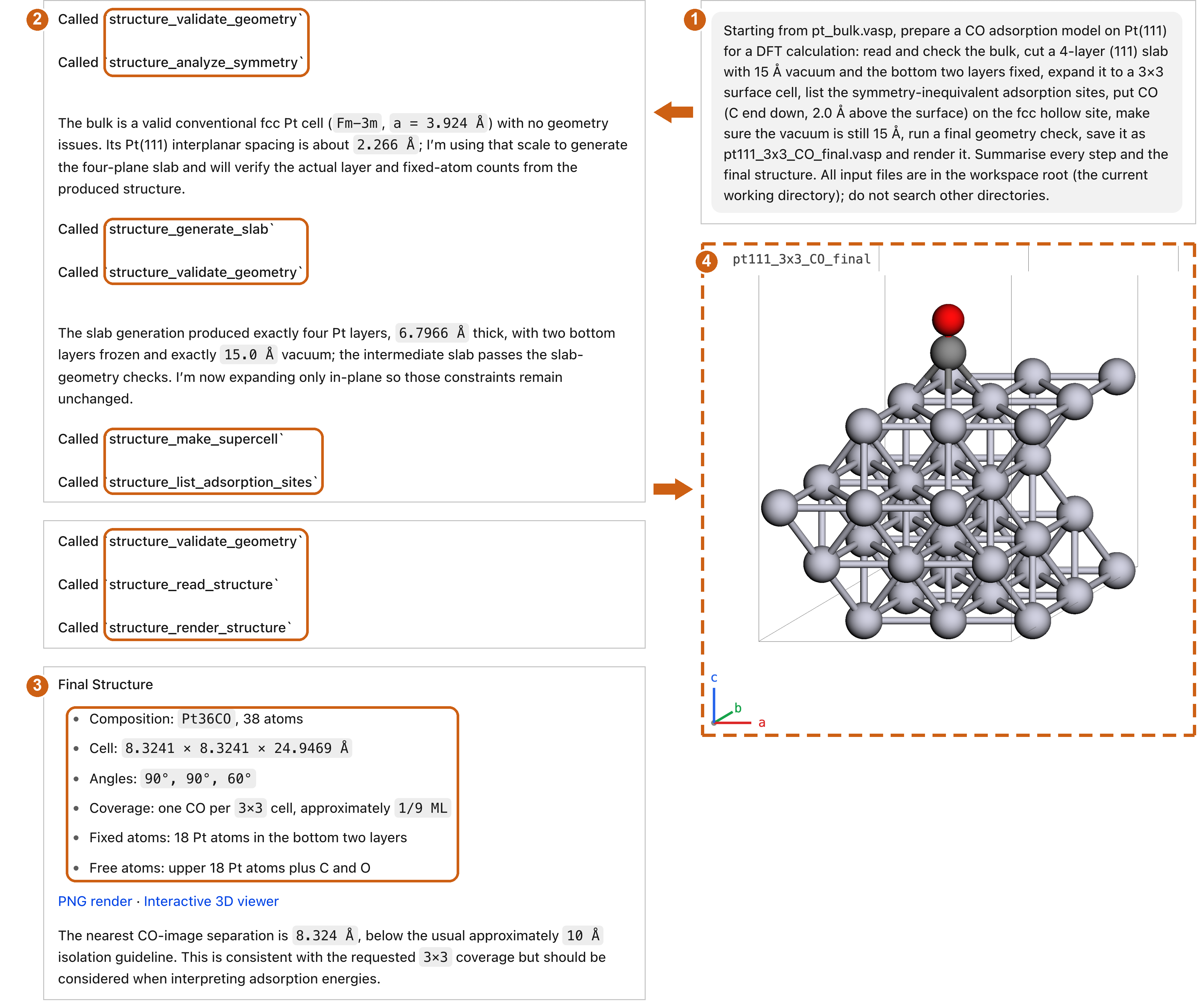}
  \caption{The complete workflow, a CO/Pt(111) adsorption model. (1) The request; (2) the intermediate steps (checking the bulk, cutting the slab, expanding it to $3\times3$ and enumerating the adsorption sites) followed by the final checks and rendering; (3) the summary of the final structure; (4) the final structure in the viewer. Tool names and parameters are boxed.}
  \label{fig:case}
\end{figure}

The request is to build a CO adsorption model on Pt(111) from bulk Pt, namely a
four-layer slab with 15~\AA{} of vacuum and the bottom two layers fixed, a
$3\times3$ surface cell and a CO molecule placed carbon end down 2.0~\AA{}
above the fcc hollow site, followed by a final geometry check, a rendering and
a summary of every step. Fig.~\ref{fig:case} shows the session.

The agent first reads the bulk structure with geometry and symmetry checking,
confirming fcc Pt ($Fm\bar{3}m$, $a = 3.924$~\AA). It then cuts a four-layer
Pt(111) slab with 15~\AA{} of vacuum and the bottom two layers fixed. The slab
thickness of 6.80~\AA{} corresponds to three interlayer spacings of
$d_{111} = a/\sqrt{3} = 2.27$~\AA. Next, the agent expands the slab to a
$3\times3$ surface cell (36 Pt atoms, 18 of them fixed), identifies the top,
bridge, hcp and fcc sites, and places CO at the fcc site specified in the
request, with a height of 2.00~\AA{} above the surface, a C--O bond length of
1.15~\AA{} and a coverage of 1/9~ML, keeping 15~\AA{} of vacuum above the
adsorbate. After the final geometry check, the summary gives the final
Pt$_{36}$CO cell ($a = b = 8.32$~\AA, $c = 24.95$~\AA) and the
selective-dynamics flags of the VASP structure format
\cite{kresse1996efficient}. The agent also gives a note that the CO molecules
in neighbouring cells are 8.32~\AA{} apart, below the roughly 10~\AA{} usually
used for an isolated adsorbate. The agent made more than ten tool calls in
total, and the researcher provided only the parameters in the request.

\FloatBarrier
\section{Conclusions}
We have presented MASON, an MCP toolkit that enables language-model agents to construct atomistic structures through dedicated tools. MASON separates workflow planning from crystallographic operations. The agent plans and verifies the workflow, while the tools execute the underlying structure manipulations. Our examples show that this framework can translate natural-language research requests into accurate and inspectable atomistic models. More broadly, the same principle can be extended to other stages of first-principles workflows, including input preparation, job execution and result analysis, for which established workflow engines \cite{huber2020aiida,ganose2025atomate2} already provide the underlying operations. Equipping agents with reliable domain-specific tools across these stages could enable increasingly autonomous first-principles research, with researchers defining scientific goals and evaluating the outcomes. MASON represents a step in this direction.
\FloatBarrier
\section*{Software availability}

MASON is open-source software under the MIT License. The source code, the user manual and ready-to-run packages for Windows, macOS and Linux (version 1.0.2) are available at \url{https://github.com/zxy-boop/MASON}. Each package contains the toolkit, its Python runtime, the agent front end and the structure viewer, and requires only access to a language model, which is configured at the first start together with an optional Materials Project API key.
\FloatBarrier
\section*{Acknowledgements}

This work was supported by the Yunnan Provincial Science and Technology United Graduate School (202402AO370001), the National Natural Science Foundation of China (Grant Nos. 12474235 and 22576107), the Key Program of Tianjin Natural Science Foundation (Grant No. 24JCZDJC01230), and the Supercomputing Center of Nankai University (NKSC).


\end{document}